\documentclass[twocolumn,superscriptaddress]{revtex4}
\usepackage[utf8]{inputenc}
\usepackage{amsmath}
\usepackage{graphicx}
\usepackage{caption}

\begin{document}

\title{Cosmic acceleration with cosmological soft phonons in a network of large structures}

\author{J. Rekier\footnote{Corresponding author}} \email{ jeremy.rekier@unamur.be}
\affiliation{Namur Center for Complex Systems (NaXys), University of Namur, Belgium}
\affiliation{FNRS Research Fellow}
\author{A. F\"uzfa} \email{jeremy.rekier@unamur.be} 
\affiliation{Namur Center for Complex Systems (NaXys), University of Namur, Belgium}

\date{\today}

\begin{abstract}
The dark energy scalar field is here presented as a mean-field effect arising from the collective motion of interacting structures on an expanding lattice. This cosmological analogue to solid-state soft phonons
in an unstable crystal network is shown to produce cosmic acceleration while mimicking phantom equation of state.  From
an analysis of the Hubble diagram of type Ia supernovae, we present constraints and discussion on the parameters of the cosmic Lagrange chain, as well as on time-variation of the soft phonon equation of state.
\end{abstract}

\maketitle
%\section{Introduction}

\emph{Introduction.} --- Thanks to many independent observations, including type-Ia supernovae Hubble diagram \cite{sn,union}, angular fluctuations of the Cosmic Microwave Background \cite{wmap} and many others based
on large-scale structures properties, Baryon acoustic oscillation \cite{bao} and galaxy redshift distortion \cite{galaxyrd}, it is now a well-established experimental fact that the expansion of the Universe is currently accelerating. What causes this cosmic acceleration is still unknown, since this observation requires modifications to either the matter-energy content of the universe and/or to the 
gravitational physics of the large-scales (through general relativity and the cosmological principle).
One can still account fairly for various observations with the help of the famous Einstein's cosmological constant $\Lambda$ but the so-called fine-tuning and coincidence problems \cite{weinberg} associated to it appear so intricate that numerous alternative explanations have arisen in the past decade. More generally, explaining cosmic acceleration without modifying gravitation requires the inclusion of a new variety of energy in the Universe, dubbed \emph{Dark Energy} (DE), whose peculiar properties modify the background cosmological expansion. The cosmological constant corresponds to perfectly frozen
DE, with absolutely no space-time variations, which is often meant as an ideal case.

Among the proposed models of DE, the hypothesis of quintessence in which DE is modeled by a scalar field provides an exciting framework\cite{Wetterich,Ratra,Copeland}. While those theories have achieved some success, they often lack to provide a sensible interpretation of the nature of this scalar field.

One of the known limitation of quintessential models arises when one wishes to interpret this new energy component in terms of fluid. To mimic the behaviour of a late time Universe dominated by a cosmological constant, one must impose $w:=p/\rho<-1/3$. Thus rendering the speed of sound in such fluid imaginary and leading to exponentially divergent instabilities at small wavelengths. This problem has already been addressed in \cite{BucherSpergel} in which the authors rightly point out that a positive speed of sound is achievable with negative pressure in a dark matter solid of sufficient shear modulus magnitude. The authors consider two possible constituents of such a dark matter solid. Namely a frustrated network of cosmic strings or domain walls. The stability of the solid prescribed in this work was later studied in \cite{Carter} in which it is found that the Dark matter solid is stable providing it is of the \emph{even type}. No inclusion of DE component is assumed in the model described by these papers as the observed acceleration of the Universe is already accounted for by the nature of the Dark Matter Solid. Neither of these papers provide any explanation as how the Dark Matter solid came to be from a randomly excited initial state.

In this work, we do not suggest any such mechanism, neither do we suggest any other candidate for the nature of the fundamental constituents of the lattice. We merely try to push the solid hypothesis further and explore how a DE scalar component may emerge from the gravitational interaction between the structures. Possibly from the kind of structure described in \cite{BucherSpergel,Carter}. This idea is most similar to the concept of phonon in solid-state physics. Indeed in material sciences, the phonon is adequately described as an effective scalar field encompassing the collective motion of atoms in a crystal. The idea of an emerging quintessence scalar field from a non-homogeneous background has already been studied in \cite{Buchert} where it is seen as coming from kinematical backreactions. We wish to suggest a different approach. 

Cosmic structure formation in an expanding universe is a very complex non-linear mechanism, in which the study of an analogue of the phonon as a mean-field behaviour for averaged quantities constitutes an intricate technical challenge. For the sake of simplicity, we rather suggest a heuristic approach by which the total Newtonian potential between the constituting structures of the lattice is limited to the first two terms in its expansion. It is later assumed that the coefficients of those terms simply rescale with cosmic expansion. This allows us to derive a simple model for a cosmological phonon as a massive scalar degree of freedom with very different cosmological dynamics than the one of quintessence. We also show that this simple model can provide cosmic acceleration through an equation of state with $w<-1/3$.

In what follows, we will build up a phonon toy-model and confront it to SNIa data before concluding on potential physical interpretations.

%\section{A toy-model of the cosmological phonon}
%\subsection{Effective lagrangian}
% AF

\emph{The Toy-Model.} ---
To write a complete model for a solid lattice Universe, one should, in principle, solve the Einstein equations with the appropriate right-hand side. This was done perturbatively in \cite{jpbjl} in the case of a cubic lattice. The observables of such a model Universe were later studied by the authors in \cite{jpbjl2}. The authors find that, in the continuous limit approximation, their model predicts a solution for the Hubble factor identical to the case of an Universe filled with dust (in the perturbative limit). 
We wish to expand on their result now by considering what happens if one includes small excitation of the lattice around its equilibrium state. 
To this end, it suffices to us to assume that the gravitational effects between the masses are adequately described by an overall Newtonian potential. Following the usual procedure in the study of large scale structures formation, one should, in principle solve the \emph{Poisson equation} $\nabla^2V=4\pi G\delta\rho$, with $V$, the gravitational potential and $\delta\rho$, the inhomogeneous part of the matter density. Because we are here only interested in the perturbation of this potential around its equilibrium value, we will assume that the potential is known in advance. It could, in principle, be derived from the solution of \cite{jpbjl} in the perfectly homogeneous case or computed in a more rigorous numerical study of large scale structure formation.

In the frame comoving with cosmic expansion, the lagrangian describing a set of masses $m_a$ undergoing long-ranged gravitational interaction mediated through the potential V is, up to quadratic terms:
\begin{equation}
L=\displaystyle\sum_a \frac{1}{2}m_a\dot{q_a}^2-\displaystyle\sum_{a,b}\frac{1}{2}k_{ab}(q_a-q_b)^2+\displaystyle\sum_a kq_a^2
\label{eq:Lsuma}
\end{equation}
where the $q_a$'s are the displacement of the masses $m_a$ from their unstable equilibrium point. $k_{ab}$ is the harmonic coupling between point masses $a$ and $b$ and $k$ is the self-coupling of a single mass:
\begin{align}
k_{ab}&=-\left(\frac{\partial^2V}{\partial q_a\partial q_b}\right)\\
k&=-\left(\frac{\partial^2V}{\partial q_a^2}\right).
\end{align} 
The function (\ref{eq:Lsuma}) can be expressed in term of a scalar function by going to the continuous limit:
\begin{align}
q_a&\rightarrow\phi(x)\label{eq:phi}\\
\displaystyle\sum_a&\rightarrow \frac{1}{l^3}\int d^3x.\notag
\end{align}
where $l$ denotes the mean distance between two structures and the integral is to be taken over the whole (possibly infinite) volume of the network. By going through the whole process, one can easily find (see e.g \cite{Zee}) that the effective lagrangian takes the form:
\begin{equation}
L=\int d^3x \frac{1}{2}\left\{\frac{1}{v^2}\left(\frac{\partial\psi}{\partial t}\right)^2-[\vec{\nabla}\psi]^2+M^2\psi^2\right\}.
\label{eq:Lcomob}
\end{equation}
The new parameters of this expression can easily be expressed in term of the network parameters:
\begin{align}
&v^2=\frac{k_{ab}l^2}{m},&M^2=2\frac{k}{k_{ab}l^2},
\label{eq:paramdef}
\end{align}
which can both be either negative or positive depending on the signs of the network parameters. For more convenience, we also have rescaled the field from (\ref{eq:phi}) as $\psi=\sqrt{\frac{k_{ab}}{l}}\phi$. The lagrangian (\ref{eq:Lcomob}) is exactly the one used in continuous treatment of elastic properties in condensed matter physics with $\psi$ being the wave amplitude of the phonon (see e.g. \cite{Simon}).

\emph{Coupling to the background.} --- We will now write this lagrangian in a form suitable for cosmology. First, we will assume that the displacement $\psi(\vec{x},t)$ is a pure scalar function such that there is no overall preferred direction for the oscillation of the field. This is in accordance with the cosmological principle that assumes isotropy of space on large scales. We thus drop the term proportional to $\vec{\nabla}\psi$ in (\ref{eq:Lcomob}). Although we do not imply that there are no difference between various polarisations of spatial displacement on small scales. Our approach is equivalent to considering only the spatially averaged scalar field $$\psi(t)=<\psi(\vec{x},t)>.$$ Where the average is taken over a region of space large enough for the cosmological principle to be vaild ($\gtrsim 100$ Mpc).
Physical quantities in the (\ref{eq:Lcomob}) will be affected by cosmic expansions through the dependency of the chain parameters in the scale factor. By looking at (\ref{eq:paramdef}) and considering the fact that the comoving distance scales as $l\sim a(t)l_0$, one can easily obtain:
\begin{align}
&\frac{1}{v^2}\sim\frac{1}{a^2},&M^2\sim\frac{1}{a^2}.
\label{eq:paramrescale}
\end{align}
% AF
This rescaling of the lagrange chain parameters will have impacts on the cosmological dynamics of the phonon, differencing it from the usual quintessence scalar field.
To write down the dynamical evolution of the Universe in presence of our new field $\psi$ we have to consider the total action consisting in the time integral of the lagrangian (\ref{eq:Lcomob}) plus the usual Einstein-Hilbert action. The (flat) Friedmann-Lema\^\i tre-Robertson-Walker (FLRW) metric being
\begin{equation}
ds^2=-N^2(t)dt^2+a^2(t)\delta_{ij}dx^idx^j,
\end{equation}
we proceed by writing the total action as the integral of an effective 1D lagrangian (see \cite{Fuzfa2006}):
\begin{equation}
L=-\frac{3}{\kappa}\frac{\dot{a}^2a}{N}+\frac{1}{2v^2}\frac{a\dot{\psi}^2}{N}-Na\frac{M^2\psi^2}{2}.
\label{eq/Leff}
\end{equation}
% AF
As a matter of comparison, the reduced lagrangian for ordinary minimally coupled scalar field $\Phi$ such as quintessence writes down
\begin{equation}
L=-\frac{3}{\kappa}\frac{\dot{a}^2a}{N}+\frac{1}{2c^2}\frac{a^3\dot{\Phi}^2}{N}-Na^3\frac{M^2\Phi^2}{2}
\end{equation}
where the quintessence potential has be taken as a mass term $V(\Phi)=\frac{M^2\Phi^2}{2}
$ for the sake of example. While quintessence scalar field comes from a fundamental field
approach, the cosmological phonon is an effective scalar field whose mass and velocity
are not constant but rather rescaled by cosmic expansion.
As a consequence, the cosmological dynamics will differ from a quintessence-filled Universe.

\emph{Cosmological dynamics} --- By varying Eq. (\ref{eq/Leff}) with respect to the generalised coordinates ($a$, $N$, $\psi$) and then specifying to the conformal gauge (N=a), we get the following equations (taking into account the contribution of matter and radiation):
\begin{align}
\frac{a'^2}{a^2}&=\frac{\kappa}{3}\left(\frac{1}{2v^2}\frac{\psi'^2}{a^2}+\frac{M^2}{2}\psi^2\right)+H_0^2\left(\frac{\Omega_m^0}{a}+\frac{\Omega_{rad}^0}{a^2}\right)\nonumber\\
\frac{a''}{a}&=\frac{\kappa}{6}M^2\psi^2+\frac{H_0^2}{2}\left(\frac{\Omega_m^0}{a}\right)\label{eq:sysdyn}\\
\psi''&=-a^2v^2M^2\nonumber\psi.
\end{align}
Where the primed quantities are to be interpreted as derivatives with respect to conformal time $\eta$ (given in terms of the synchronous time by $dt=a(t)d\eta$) and with $\kappa=8\pi G$ in units with $c=1$. The zero-indiced symbols denote the present values of these quantities (we took $\Omega_m^0=0.25$, $\Omega_{rad}^0=7.97\times 10^{-5}$, $H_0=71$km/s/Mpc)\cite{union,wmap}. Defining
\begin{equation}
\Omega_\psi(a)=\frac{\kappa}{3 a^2H^2}\left(\frac{1}{2v^2}\frac{\psi'^2}{a^2}+\frac{M^2}{2}\psi^2\right),
\end{equation}
the Friedmann equation can be put under the form:
\begin{equation}
\frac{a'^2}{a^2}=\frac{H_0^2}{(1-\Omega_\psi(a))}\left(\frac{\Omega_m^0}{a}+\frac{\Omega_{rad}^0}{a^2}\right).
\label{eq:Hconstraint}
\end{equation}

To get insight about which values the parameters $v^2$ and $M^2$ should take, one may look at the asymptotic behaviour of the model. 

From the cosmological equations ($\ref{eq:sysdyn}$) written in the synchronous gauge one can easily guess the ratio of the pressure over the energy density of the field $\psi$ to be 
\begin{equation}
w_\psi=\frac{1}{3}\frac{\left(\frac{\dot{\psi}^2}{v^2a^2}-M^2\psi^2\right)}{\left(\frac{\dot{\psi}^2}{v^2a^2}+M^2\psi^2\right)}.
\label{eq:wpsi} 
\end{equation}
% AF
This equation of state for the effective field $\psi$ has interesting properties. Indeed, it interpolates between $w_\psi=-1/3$ when the field is frozen ($\frac{\dot{\psi}^2}{v^2a^2}\ll M^2\psi^2$) and a cosmological constant $w_\psi=-1$ in the far future ($a \gg 1$ ; de Sitter Universe). To show this is an asymptotic solution, we assume that the Universe enters is acceleration regime around $a\sim1$ and start from the condition $w_\psi=-1$, which can be readily integrated for the asymptotic behaviour of $\psi$ from (\ref{eq:wpsi}) as
\begin{equation}
\psi=\psi_0\exp \left(\sqrt{|\frac{v^2M^2}{2}|}(t-t_0)\right),
\label{eq:psipsi0}
\end{equation}
with $\psi_0$ the present-day value of the field. 
Inserting this result into the acceleration equation and solving for $a$, we find 
\begin{equation}
a=\sqrt{\frac{\kappa}{6}}\frac{\psi_0}{\sqrt{|v^2|}}\exp\left(\sqrt{\left|\frac{v^2M^2}{2}\right|}(t-t_0)\right).
\label{eq:adesitter}
\end{equation}
% AF
By a suitable choice of $\psi_0$ and $\sqrt{|v^2|}$, one can always make $\sqrt{\frac{\kappa}{6}}\frac{\psi_0}{\sqrt{|v^2|}} \approx a_0$. The result (\ref{eq:adesitter}) is then to be compared with the usual de Sitter asymptotic solution $a=a_0\exp\left(\sqrt{\frac{\Lambda}{3}}c(t-t_0)\right)$, showing that the asymptotic state is a de Sitter universe (see also Fig. 1).

Therefore, the cosmological dynamics of the phonon effective field departs from that of a massive scalar field (quintessence with quadratic potential) for which 
$w\rightarrow -1$ when the field is frozen and $w$ being zero on average (and the field behaving as pressureless matter) asymptotically. Note that the fact that the value of $w_\psi=-1/3$ when the field is frozen features an interesting connexion to the work of \cite{BucherSpergel} concerned with solid Dark matter made of cosmic strings in equilibrium which, in principle, prevents one to distinguish between both based solely on cosmological dynamical grounds.

\emph{Constraints from experiments.} --- The problem of cosmic coincidence is not well addressed in this work since we had to tune the value of the parameters to produce an acceleration phase triggered around $a\sim1$. However, by pressing forward we shall see if the model predicts interesting results for the dynamics of the Universe between the time of formation of the cosmological lattice and the present epoch.
From (\ref{eq:adesitter}), it is straightforward to identify the quantity representing an effective cosmological constant to be:
\begin{equation}
\Lambda_{\text{eff}}=\frac{3}{2}\frac{|v^2M^2|}{c^2}.
\label{eq:Lambeff}
\end{equation} 
That is, if we fit the value of the cosmological constant to the Supernovae data, then any two values of the parameters $v^2$ and $M^2$ satisfying the constraint (\ref{eq:Lambeff}) should produce the same asymptotic behaviour. 

% AF
We found that the parameter $v^2$ has to be negative for the asymptotic solution of (\ref{eq:sysdyn}) to be de Sitter (hence the need for the absolute value in (\ref{eq:psipsi0})). 
The physical meaning of this is understood by looking at the third equation of (\ref{eq:sysdyn}). %
Positive values of $v^2$ would imply oscillations of the field through time while negative values describe an inelastic deformation. The latter case is more suited to describe the gravitational instabilities in the interplay of the large structures in a dynamical Universe (after BAO epoch). 
% AF
This case of facts is to be understood as the analogue of the concept of \textit{soft} phonon of solid state physics (see e.g. \cite{Kittel}), which represent fastly growing inelastic deformations when a crystal departs from an unstable equilibrium state. This process is responsible for structural phase transitions appearing when the frequency of the phonon becomes imaginary, thus leading to instabilities.
%
%In the next section, we analyse the dynamical behaviour of these equations.
%\begin{section}{Dynamical analysis}

% AF
The dynamics of the system (\ref{eq:sysdyn}) is investigated numerically. We start the integration at $a_i=10^{-3}$, corresponding to the beginning of the process of large-scale structure formation through gravitational collapse. Before that time, acoustic oscillations in the primordial plasma prevent inelastic deformation in the network. We also had to impose initial conditions on the values of $\Omega_\psi^i$ and $\psi'_i$ at this very point. While we chose to start at rest $\psi'_i \approx 0$, $\Omega_\psi^i$ is a free parameter found to be less than 1\% (see below).

We chose to integrate the second and the third equation of (\ref{eq:sysdyn}) and then check the solution by injecting the result in the Friedmann equation (\ref{eq:Hconstraint}). The constraint is always verified up to the machine accuracy. The integration has been carried out for several values of $v^2$ and $M^2$. The model fits the SNIa data\cite{union} with $\bar{\chi^2}=1.0018$ (in the simile, $\bar{\chi^2}_{\Lambda\text{CDM}}=0.99$ for the same data set).

Fig. \ref{fig:qa} shows the compared evolution of the acceleration parameter ($q=\frac{\ddot{a}a}{\dot{a}^2}$).  The phonon model predicts a stronger yet smoother acceleration than $\Lambda$CDM at all time. This should have impacts on the physics of large-scale structure formations.

\begin{center}
	\includegraphics[width=0.45\textwidth]{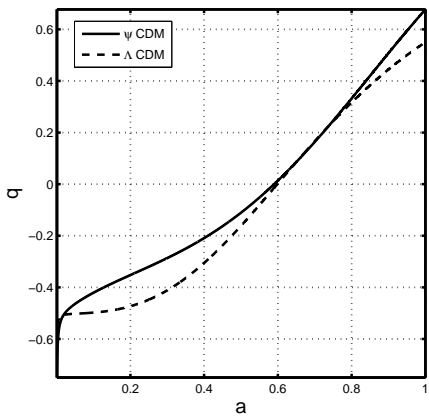}
	\captionof{figure}{Evolution of the acceleration parameter ($q=\frac{\ddot{a}a}{\dot{a}^2}$) for the cosmological phonon model ($\psi$ CDM) as compared to $\Lambda$CDM.}
	\label{fig:qa}
\end{center}

Another departure from $\Lambda$CDM lies in the evolution of the equation of state parameter $w_\psi$ shown on fig. \ref{fig:wpsia}. As mentioned above, this shows that the energy density of the phonon interpolates between a Nambu-Goto string gas ($w_\psi=-1/3$) at early times and a cosmological constant asymptotically. This model evolves with a phantom equation of state ($w<-1$) around today without any appeal to any non-minimal coupling of gravity.
\begin{center}
	\includegraphics[width=0.45\textwidth]{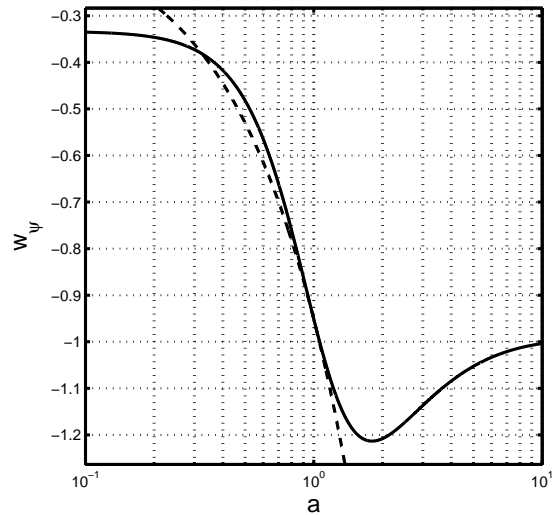}
	\captionof{figure}{Evolution of $w_\psi$ from the time of recombination to the far future.}
	\label{fig:wpsia}
\end{center}

The dotted line of fig. \ref{fig:wpsia} shows the linear approximation of the curve around $a=1$ : $w_\psi(a)=w_0+w_a(1-a)$ \cite{wowa} with $w_0=w_\psi(a_0)$. Fig. \ref{fig:om0_w0} shows the confidence regions at $1\sigma$ and $2\sigma$ level in the plane of parameters ($w_0$,$\Omega_m^0$). We see that, amongst the compatible models, one recovers a model similar to $\Lambda$CDM ($w_0=-1$) at the limit of $1\sigma$ level only for lower values of $\Omega_m^0$.% with a slightly bigger age of the Universe (15Gyr). %The lines represent the models giving an age (in Gyr) for the Universe specified by the corresponding label.

\begin{center}
	\includegraphics[width=0.4\textwidth]{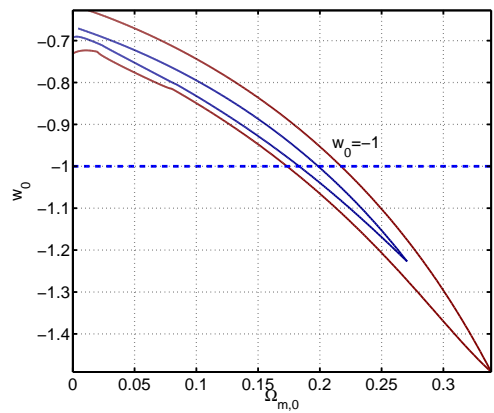}
	\captionof{figure}{$1\sigma$ and $2\sigma$ level confidence regions in the plane of parameters ($w_0$,$\Omega_m^0$) (data from \cite{union}) ($H_0=71$km/s/Mpc)} %. The straight lines collect the models with a given age of the Universe
	\label{fig:om0_w0}
\end{center}

\begin{table}[h]
\centering
\begin{tabular}{| c | c |}
	\hline
	Parameter & best-fit $\pm 1\sigma$\\ \hline
	$\Omega_m^0$ & $0.14_{-0.14}^{+0.13}$\\ \hline
	$\Omega_{\psi}^0$ & $0.86_{-0.13}^{+0.14}$\\ \hline
	$q_0$ & $0.66_{-0.14}^{+0.19}$ \\ \hline
	$w_0$ & $-0.90_{-0.33}^{+0.22}$ \\ \hline
	$w_a$ & $0.77_{-0.31}^{+0.36}$\\ \hline
\end{tabular}
\caption{Best-fit of the cosmological parameters}
\label{teb:tabatiere}
\end{table}

The best-fitted values of the various cosmological parameters and the boundaries of the first confidence region ($1\sigma$ level) are summed up in Table 1. The best-fit value for $\Omega_m^0$ is small compared to the estimation from \cite{union}. Still the usual constraints on $\Omega_m^0$ of order $0.25$ are compatible with other predictions to $1\sigma$ level (see e.g. \cite{om0cons}) .

\emph{Discussion.} --- The preceding results show that the phonon model could be a nice alternative to $\Lambda$CDM with no more than two free parameters: the initial amount of energy in $\psi$ and the self-coupling parameter $M^2$ (the $v^2$ parameter being constrained by (\ref{eq:Lambeff})). It reproduces a de Sitter Universe for $a\gg1$ and provides an interesting interpretation of the nature of DE without involving any speculative physics. This work starts from the considerations of \cite{jpbjl,jpbjl2} and \cite{BucherSpergel,Carter} by first considering a Universe filled with a lattice of large scale structures and then investigating the effects of phonons propagating on this lattice. One of the crucial features of the latter works was to make sure that the network thus created was stable so that the deformation of the lattice is only elastic. We fail to see how large structures inside the same particle horizon should form a stable lattice similar to those found in solid state physics. External perturbations aside, the latter are a stable configuration of atoms as the repulsive forces between the ions prevent them to collapse on one another. As for the case of large structures, the dominant force between being the attractive gravitational force would, contrarily, induces collapse of the lattice cells except if the repulsive effect of expansion is already large enough to compensate the local gravitational interactions. 
The present work circumvent this problem by assuming that instabilities are indeed present and play a crucial role in explaining the late-time acceleration of the Universe since the phonons in question here are soft. It all appears as if the lattice experiences a phase transition during its history. This would in turn tend to explain the strange variation of $w_\psi$ predicted over time. A direct test of the present model could be provided by direct measurements of these variation.

We would like to suggest the following history of the Universe after recombination:
First, the structures start to form thus creating a lattice of small masses possibly adequately described by the work of \cite{BucherSpergel,Carter}. As the average mass of the structures gets bigger, the lattice experiences a phase transition due to gravitational attraction between the structures. This leads to instabilities in the phonon field that can drive the start of the accelerated expansion as shown in the present work. The repulsive effect of the acceleration of the Universe then prevents the occurrence of another phase transition. 
However, this model does not address the coincidence problem has it involves some tuning of the cosmic chain parameters on the value of the measured cosmological constant to ensure the acceleration occurs around the present epoch and so does not explain why the phase transition should have happened at the time predicted by the model. However, the need for cosmic structures as a medium for the phonon prevents the acceleration to happen during the radiation dominated era.

In any case, it is likely that the aforementioned instabilities should have laid imprints on large angular scales of the CMB sky. Future directions would then include a detailed study of the implication of the cosmological phonon on the CMB moments. Another important issue would be to derive the preceding model from conventional cosmological perturbation and statistical analysis in inhomogeneous cosmology. Challenges for numerical N-body simulations of large-scale structure formation would be to estimate the network parameters from the overall potential energy generated by the inhomogeneous density field. Once this has been done, one can then check whether they can fall in the range predicted by the SNe Ia analysis. Such numerical studies will also allow one to compute the structure network potential. Another interesting question to be addressed by  N-body simulations would be to investigate whether collective motions corresponding to the propagation of a soft phonon on large scales can be reproduced. 

Further theoretical considerations would include higher order-coupling of Dark Matter to the phonon field or other solid-state phonon-inspired observables, as well as a complete study of the instability wavefront and large-scales collective motion.
We also reckon a treatment somewhat more rigorous than the heuristic approach developed here would be needed if one wishes to investigate the issues pointed out here in details. Solace might be found by going back to \cite{jpbjl} and investigate how our soft phonon description might emerge from the introduction of some defects or perturbations in this model. This further work might provide the tensorial description that our present model is currently lacking.
\section*{Acknowledgements}
The authors warmly thank L.Henrard for the useful discussions from which this paper originates. J.Rekier is supported by an FRS-FNRS (Belgian Fund for Scientific Research) Research Fellowship. This research used ressources of the 'plate forme technologique en calcul intensif (PTCI)' of the University of Namur, Belgium, for which we acknowledge the financial support of the F.R.S.-FNRS (convention No. 2.4617.07. and 2.5020.11)

%\end{section}

\end{document}